\documentclass[]{revtex4}
\begin{document}
\title{RENORMGROUP INVARIANTS AND APPROXIMATIONS OF MAPPINGS}
\author {Gennady N. Nikolaev\\{\it Institute of  Automation and Electrometry of SB RAS}, \\ \it{ Pr. Koptyuga 1, Novosibirsk, 630090,
Russia}}
\date {November 18, 2004}
\begin{abstract}
The relationship between mappings of sets and renormalization
group transformations is established, and renormalization group
invariants of such mappings are found. These results are valid
both for continuous and discrete mappings and for various
dimensionality of image and preimage of the mappings too. It is
suggested a number of mapping approximations improved in
comparison with an ordinary power expansion. The approximations
take  into account the global one-to-one character of the
mappings. The method is illustrated by a number of examples: by
reconstructing of some analytical functions, calculating the
integral of the typical partition function of statistical
mechanics and the ground state energy for the quartic anharmonic
oscillator. In the whole range of nonlinearity parameter varying
from zero up to infinity the accuracy of the RG approximation
based on a few terms of divergent series is about 0.06\% in the
next to last case and 0.004\% in the last  one.

\end{abstract}
\maketitle
\section{Introduction}\label{c:int}

Renormalization group (RG) conception originally arose in the
beginning of the fifties \cite{Stuckelberg53} as a result of the
discovery of special group of continuous transformations in quantum electrodynamics. This transformations are connected with a complicated procedure of renormalization, that is `removing of ultra-violet infinities'. It turned out, that physical quantities do not vary at simultaneous rescaling of the 4-momentum transfer squared $z$, mass squared $y$, and also special transformation of the dimensionless charge squared $g \; (g \rightarrow \overline{g}(z,y,g))$.  The invariant charge $\overline{g}(z,y,g)$  obeys to the equation \cite{Gell-Mann54,Bogoliubov55}
\begin{equation}\label{e1}
\overline{g}(z,y,g)=\overline{g}(z/{t},y/{t},\overline{g}({t},y,g)).
\end{equation}
For the massless quantum-field model instead of (\ref{e1}) the following relation is valid
\begin{equation}\label{e2}
    \overline{g}(z,g)=\overline{g}(z/{t},\overline{g}({t},g)).
\end{equation}
The generalization of equation (\ref{e2}) for the two-charge
quantum-field model was proposed also \cite{Bogoliubov55}. In
this case, instead of expression (\ref{e2}) the following two
relations are fulfilled
\begin{equation}\label{e3}
\overline{g}_{i}(z,g_1,g_2)=\overline{g}_i(z/{t}, \overline{g}_{1}(t,g_1,g_2),\overline{g}_{2}(t,g_1,g_2))
\qquad i=1,2
\end{equation}
The general solution of such functional equations was obtained \cite{Ovsyannikov56}.

The rapid expansion of the RG method far beyond the quantum-field
theories began since the seventies. It was strongly promoted by
the article \cite{Wilson71}. In this article the conception of an
approximate RG  was introduced for analysis of thermodynamic
systems behaviour near the phase transition points. Afterwards
the RG method was applied for description of turbulence
\cite{Dominics79,Pelletier80}, polymeric compounds
\cite{Gennes79}, radiation transfer in opaque mediums with the
strong frequency dependence of the quantum path length
\cite{Bell78}, fractals \cite{Suzuki83}, scripts of the dynamic
systems transitions to the determined chaos \cite{Schuster84},
and also was applied to other problem. The relation was found out
between nonlinear problems of radiation transport and additive
variant of the RG \cite{Mnazakanyan78,Mnazakanyan82}. This result
was extended on the wide range of the physical problems; and the
functional self-similarity concept, that generalizes the usual
self-similarity concept, was introduced \cite{Shirkov82}. The
modern state of the researches concerned with use of this concept
for boundary-value problems of the mathematical physics is
represented in the recent review \cite{Kovalev99}.

    In the present paper a relationship between mappings from one set into another one and RG transformations is established, and RG invariants of such mappings are found. On this basis a number of RG approximations  that are founded on a few terms of series of one-to-one functions is constructed. They represent better approximations  of the sought function in comparison with its Taylor series expansion because the RG-approximations inherits its one-to-one nature.

    The necessity for reasonable approximations of physical quantities is connected to  the fact that the majority of realistic problems cannot be solved exactly. To solve them, one uses various approximating methods that are in general different kinds of perturbation theories near a chosen zero approximation. These methods are evidently convenient if the perturbation (interaction or coupling) parameter  $g$  is small, $g \ll 1$.  However, if $g \gtrsim 1$ the expansion in powers of $g$ have no sense. But for many realistic systems, because of their complexity, it is technically impossible to use other calculation techniques except for the perturbation one. In such a case there are a number of resummation (or reconstruction) methods that allow to find the value of some function $f(g)$ in the case of $g \gtrsim 1$ using only the results of its calculation in the limit $g \ll 1$. These methods include the improved perturbation theory, the Pad\'{e} approximation, the Borel summation, the conformal mapping and their combinations \cite{KazaShirk80,Zinn81}, the continued fraction approximations \cite{JonesThron80}, and the Romberg algorithm \cite{Beleznay86}. These methods, to be accurate enough, need the calculation of many subsequent approximations for the sought function. But if system is so complex that  we are able to calculate solely a couple of approximations, then the majority of these methods lose their applicability. In this situation the method of RG approximation  seems especially relevant because of its  quite good accuracy when using only a few terms of the perturbation theory.

\section{Renormgroup invariants of one-to-one mappings}\label{c:1to1 maps}

At first let us consider a function of a single variable $f(x)$
as a primary mapping. It is assumed hereinafter that the function
$f(x)$  is one-to-one, i.e. it has the inverse function. Now we
consider the value of this function at the argument $x_1+x$.
Let us express the quantity $x$  via the inverse function:
$x=f^{-1}(f(x))$.  This relation allows one to write down
$f(x_1+x)$ as follows:
\begin{equation}\label{e4}
  f(x_1+x)=f(x_1+f^{-1}(f(x))) \equiv F(x_1;f(x)).
\end{equation}
Thus, we have presented $f(x_1+x)$  as a new function $F$  of
two variables, value of the initial function $f$  at the
original argument $x$  and the translation $x_1$.  According
to definition (\ref{e4}), the value of the function $F$  at
$x_1=0$  coincides with the value of the initial function $f$  at
the argument $x$:
\begin{equation}\label{e5}
  F(0;f(x))=f(x)
\end{equation}

Now let us consider the value of the function $f$  at the
argument
$\tilde{x}=x+x_1+x_2$  displaced sequentially on $x_1$  and
$x_2$ with respect to $x$.  The value of the function
$f(\tilde{x})$ can be represented via the function $F$ by two equivalent
ways.
According to the one way, it is possible to consider the two
sequential displacement $x_1$ and $x_2$ as a whole
displacement $x_w=x_1+x_2$ and to take advantage of the
definition (\ref{e4}), that leads to the relation
\begin{equation}\label{e6}
  f(x+(x_1+x_2))=F(x;f(x_1+x_2)).
\end{equation}
And then it is possible to take advantage of definition
(\ref{e4}) once more to transform the function $f(x+x_2)$ on the
right-hand side of expression (\ref{e6}) in terms of
$F$:
\begin{equation}\label{e6a}
  F(x;f(x_1+x_2))=F(x;F(x_1;f(x_2))).
\end{equation}

On the other way, at first one can use definition (\ref{e4}) with
replacement $x+x_1$ for $x$, that results in
\begin{equation}\label{e7}
  f((x+x_1)+x_2)=F(x+x_1;f(x_2)).
\end{equation}

Equating the right parts of expressions (\ref{e6a}) and
(\ref{e7}), taking into account equality (\ref{e6}), we
get the functional equation
\begin{equation}\label{e8}
  F(x+x_1;f))=F(x;F(x_1;f)).
\end{equation}
Hereinafter we have omitted for brevity the argument $x$ of the
function $f$. Equation (\ref{e8}) coincides in its structure
with the equation (\ref{e1}) for the radiation transport in  \cite{Mnazakanyan82}.

Note that relation (\ref{e8}) turns into (\ref{e2}) by means of the
replacements $x=\ln(z/t)$, $x_1=\ln(t)$, and the obvious
change of the notations: $f \rightarrow{g}$, $F
\rightarrow{\overline{g}}$.

From equation (\ref{e8}) the partial differential equation for
the function $F$  is easily obtained
\cite{Mnazakanyan82,Shirkov82,Kovalev99}. For this purpose it is
enough to differentiate both parts of equation (\ref{e8}) on $x_1
\rightarrow 0$ and to take advantage of property (\ref{e5}):
\begin{equation}\label{e9}
  [\frac{\partial}{\partial x}-\beta(f) \frac{\partial}{\partial f}]F(x;f) \equiv R F(x;f)=0;
\end{equation}
\begin{equation}\label{e10}
  \beta (f)=\frac{\partial F(x_1;f)}{\partial x_1}|_{x_1=0}.
\end{equation}

Functional equation (\ref{e8}) expresses the invariance of $F$
under the RG transformation \cite{Shirkov82,Kovalev99}
representing a simultaneous one-parametric point transformation
of both arguments of the function $F$:
\begin{equation}\label{e11}
  T(x_1): \qquad {x \rightarrow{\tilde{x}}=-x_1+x, \qquad f \rightarrow {\tilde{f}}= F(x_1;f)}; \qquad F(0;f)=f.
\end{equation}
Equation (\ref{e8}) guarantees fulfillment of the group property
$T(x_1)T(x_2)=T(x_1+x_2)$. Note that differential equation
(\ref{e9}) represents the infinitesimal form of transformation
(\ref{e11}), and $R$ is the infinitesimal operator of
RG-symmetry.

\section{Renormgroup approximations of one-to-one mappings}\label{c:1to1 apprx}

Frequently it is possible to define the behavior of some sought
function only in a vicinity of some point. Such situation is
typical for iterative procedures or for one of the basic
methods of theoretical and mathematical physics, perturbation
theory. As a rule, the calculation of the higher-order
corrections is connected with significant difficulties of the
technical or basic character (for example, because of the
divergence of the higher-order corrections). It is known also
that a finite series function approximation often describes
wrongly the long range function behaviour. Relations (\ref{e8}),
(\ref{e9}) allow to receive improved approximation of the sought
function (so-called RG method \cite{Bogoliubov55}). This
approximation more correctly describes the long range behavior of
the sought functions. The solution of equation (\ref{e9}) with
boundary condition (\ref{e5}) is possible to represent as
\begin{equation}\label{e12}
  F(x;f)=X^{-1}(x+X(f)),
\end{equation}
\begin{equation}\label{e13}
  X(f)=\int{\frac{df}{\beta (f)}} \, ,
\end{equation}
where $X^{-1}$ is inverse to $X$ function. Ordinarily, to
determine the function (\ref{e13}) one use an approximate value
for $\beta$ (\ref{e10}) calculated in the small parameter limit
$f=0$ . Let us consider other approach that does not presuppose
the smallness. According to definitions (\ref{e10}) and (\ref{e5})
we have:
\begin{equation}\label{e14}
  \beta (f)=\frac{df}{dx} \equiv f^{'}(x).
\end{equation}
Suppose that we know an approximate expression of $f(x)$ in a
neighborhood of some point $x_3$, for example, as a Taylor
series  (without loosing of generality one can set $x_0=0$):
\begin{equation}\label{e15}
  f(x)=a_0+a_{1}x+a_{2}x^2+a_{3}x^3+ \ldots .
\end{equation}
Series (\ref{e15}) can be inverted:
\begin{equation}\label{e16}
  x=\phi + b_2{\phi}^2+b_3{\phi}^3+ \ldots,
\end{equation}
where $\phi=(f-a_0)/a_1$, the coefficients $b_i$ are expressed
via the coefficients of initial series (\ref{e15}). The sequential
substitution of expansion (\ref{e15}) and (\ref{e16}) into the
expression (\ref{e14}) enables us to get the required dependence
of $\beta(f)$ as an infinite series on $\phi$. If we use
various finite series approximations of $\beta (f)$ , we receive
various RG approximations of the sought function by the inversion
of expression (\ref{e13}). So, the linear approach of $\beta
(f)$ gives the following RG approximation of the sought function
$f(x)$:
\begin{equation}\label{e17}
  f(x) \simeq f^{(2)}_{\beta}(x)=a_0+\frac{a_1{^2}}{2a_2}\left[\exp(2a_{2}x/a_1)-1\right].
\end{equation}
From the square-law approach of $\beta (f)$ another
RG-approximation follows:
\begin{equation}\label{e18}
  f(x) \simeq f^{(3)}_{\beta}(x)=a_1+\frac{a_1{^2}}{a_{2}}\left[\frac{a_1}{a_{2}}\kappa \coth(\kappa x)-1\right]^{-1},
\end{equation}
where $\kappa=\sqrt{3[(a_2/a_1)^2-a_3/a_1]}$.

It is interesting to note, that expression (\ref{e13}) represents
exactly the variable $x$  as a function of $f$ taking into
account (\ref{e14}). Thus, expressions (\ref{e13}) and
(\ref{e16}) are equivalent. Therefore, one can get one more
variant of the RG function $f$  approximation, using only a few
sequential terms of series (\ref{e16}) and inverting the obtained
approximate equation with respect to $f$. So, the square-law approach of series (\ref{e16}) results in  the following RG approximation:
\begin{equation}\label{e19}
   f(x) \simeq f^{(2)}_{X}(x)=a_0+\frac{a_1{^2}}{2a_2}[1-\sqrt{1-4(a_{2}/a_1)x}]
  =a_0+\frac{2a_{1}x}{1+\sqrt{1-4(a_{2}/a_1)x}}.
\end{equation}

It is possible also to receive other variants of the function RG
approximations, using other approaches for $f(x)$, $f^{'}(x),$
or $x(f)$  instead of the Taylor series . For example, one can
get another RG approximation of the sought function using second
convergent of the continued fraction corresponding to series
(\ref{e16}):
\begin{equation}\label{e20}
  f(x) \simeq f^{(2)}_{Xcf}(x)=a_0+\frac{a_{1}x}{1-(a_{2}/a_1)x}
\end{equation}
It should be noted that this expression coincides with the
approximate one obtained from (\ref{e19}) as a result of the
square root expansion. It is interesting to note  also the coincidence of this expression with the self-similar approximation \cite{Yukalov90}.

Let us represent the estimations of the RG approximations accuracy of some functions for $x=1$. We use the notation $\Delta_{a}=[f_{a}(1)-f(1)]/f(1)\times 100\%$, where $f_a(x)$  is any of the RG approximations of a function $f(x)$, as a measure of accuracy.  For the function $f(x)=\ln{(1+x)}$  we have
$\Delta_{\beta}^{(2)} \approx -8.8$, $\Delta_{\beta}^{(3)}
\approx 1.6$, $\Delta_{X}^{(2)} \approx 5.3$,
$\Delta_{Xcf}^{(2)} \approx -3.8$, $\Delta_{T}^{(2)} \approx
-27.8$, $\Delta_{T}^{(3)} \approx 20.2$, where the last two quantities refer to the finite Taylor series approximations of the square-law type and cubic one respectively. The  accuracy of the different RG approximations for other function $f(x)=x/\sqrt{1+x}$ is represented as follows: $\Delta_{\beta}^{(2)} \approx -10.3$, $\Delta_{\beta}^{(3)}
\approx 3.4$, $\Delta_{X}^{(2)} \approx 3.5$,
$\Delta_{Xcf}^{(2)} \approx -5.4$, $\Delta_{T}^{(2)} \approx
-28.9$, $\Delta_{T}^{(3)} \approx 24.0$.

It should be noted that the function RG approximations for
$a_1=0$  can be obtained in the same way. In this case $x(f)$  is
represented as a power series on $\phi_1=\sqrt{(f-a_0)/a_2}$.

   It is important to note, that the expansion in powers of $x$ of both sought function $f(x)$ and its RG-approximation $f_{a}^{(n)}(x)$ coincide up to the order $n$ inclusive. Therefore a successive RG-approximations $f_{a}^{(n)}(x)$ of the sought function $f(x)$  converge to the latter one  not worse of the Taylor series while value $n$ sequentially increases. RG-approximations of one-to-one function represent in general its better approximations in comparison with its Taylor series expansion because  RG-approximations inherits its one-to-one nature.

\section{Partition functions of statistical mechanics}\label{c:PF}

Let us illustrate accuracy of our method on the two important physical examples. Both of them satisfy the following conditions: (a) the standard perturbation theory is not applicable at all leading to divergent series; (b) some analytical expressions are known; (c) analytical or numerical calculations are available giving the possibility to compare them with obtained results; and (d) there are other approximations that may also be compared with the method considered.

    First example is an integral having the mathematical structure typical for the partition functions for statistical problems with effective potential of the form $V(\phi)=\phi^2+g \phi^4$, that is, for the problems with the so-called $\phi^4$ interaction. Consider the integral

\begin{equation}\label{e30}
  I(g)=\frac{1}{\sqrt{\pi}} \int_{-\infty}^{\infty} {\exp(-\phi^2-g \phi^4) \, d\phi}
\end{equation}
with $g \in (0,\infty)$. The generating functional of the $\phi^4$ quantum field theory has also the structure of (\ref{e30}).

Integral (\ref{e30}) admits an exact quadrature expressed via the modified Bessel function of the second kind:
\begin{equation}\label{e31}
  I(g)=\frac{\exp(1/{8g})}{\sqrt{4\pi g}}K_{1/4}(\frac{1}{8g}).
\end{equation}
    It is easy to find the expansion of (\ref{e30}) in powers of $g$:
\begin{equation}\label{e32}
  I(g)=\sum \limits_{n=0}^{\infty}\frac{(-1)^n}{\sqrt{\pi} n!} \Gamma \left(2 n +\frac{1}{2}\right)\, g^n.
\end{equation}
The strong coupling limit $g \gg 1$ of (\ref{e30}) can be obtain by using saddle-point method
\begin{equation}\label{e32b}
  I(g) \simeq \frac{1}{2 \sqrt{\pi}} \left[\Gamma \left(\frac{1}{4}\right)g^{-1/4}-\Gamma \left(\frac{1}{4}\right)g^{-3/4}\right]; \qquad  g \gg 1 \, . \end{equation}
Substituting here the quantities $\Gamma \left( 1/4 \right) \approx 3.624\,409$, $\Gamma \left( 3/4 \right)\approx 1.225\,417$ for gamma-function $\Gamma(y)$, we get
\begin{equation}\label{e32c}
  I(g) \simeq 1.022\,735\, g^{-1/4}-0.345\,384\, g^{-3/4} \, ; \qquad  g \gg 1 \, .
\end{equation}

    As can see, the $n$th term in the expression (\ref{e32})  diverges at any non-zero $g$, as $n \to \infty$, since
\begin{equation}\label{e33}
  \frac{\Gamma (2 n+1/2)}{n!} \simeq \left(\frac{4n}{e}\right)^n \rightarrow \infty \, , \qquad  n \rightarrow \infty \, .
\end{equation}
So, expansion in powers of $g$ has no sense at $g\sim 1$. Retaining three terms of series (\ref{e32}), we get the approximation of (\ref{e30})
\begin{equation}\label{e34}
  I(g)\simeq 1-\frac{3}{4}g+\frac{105}{32}g^2 \, ; \qquad  g \ll 1 \, .
\end{equation}
The accuracy $\Delta_{T}^{(2)}$ of the approximation (\ref{e34}) is about 360 \% at $g=1$. On the other hand,  we have $\Delta_{\beta}^{(2)} \approx -18\%$, $\Delta_{X}^{(2)} \approx 7\%$, and $\Delta_{Xcf}^{(2)} \approx 9\%$ for RG approximations (\ref{e17}), (\ref{e19}), and (\ref{e20}) respectively that are based on approximation (\ref{e34}).  It is seems satisfactory accuracy for approximations based on a few terms of divergent series.

\subsection{Modified RG approximation.}\label{c:M RG PF}

One can get more precise approximation by our method combined with the principle of minimal sensitivity (PMS) \cite{Stev81} even in the whole range $g \in (0,\infty)$.

    Let us rewrite partition integral (\ref{e30}) in the equivalent form
\begin{equation}\label{e35}
  I(g)=\frac{1}{\sqrt{\pi}} \int_{-\infty}^{\infty} {\exp(-z^2\phi^2+\triangle V) \, d\phi} ,
\end{equation}
where
\begin{equation}\label{e36}
   \triangle V \equiv (z^2-1)\phi^2-g\phi^4,
\end{equation}
and $z$ is trial parameter. Expanding the integrand of (\ref{e35}) in powers of (\ref{e36}), we get
\begin{eqnarray}\label{e37}
  I_{1}(g,z)&=&\frac{1}{z} \left[1+\frac{1}{2}\left(1-\frac{1}{z^2}\right)-\frac{3g}{4z^4}\right] ,\\
  I_{2}(g,z)&=&I_{1}(g,z)+\frac{1}{z} \left[\frac{3}{8}\left(1-\frac{1}{z^2}\right)^2-\frac{15g}{8z^4} \left(1-\frac{1}{z^2}\right) + \frac{105g^2}{32z^8}\right] .\label{e38}
\end{eqnarray}
in the first and in the second order respectively.

Let us make the substitution $x=1-1/z^2$. Then expressions (\ref{e37}) and (\ref{e38}) transform into
\begin{eqnarray}\label{e39}
  I_{1}(g,x)&=&\sqrt{1-x} \left[ 1 + \frac{1}{2}x - \frac{3}{4}g(1-x)^2\right],\\
  I_{2}(g,x)&=&I_{1}(g,x)+\sqrt{1-x} \left[\frac{3}{8}x^2-\frac{15g}{8}x(1-x)^2 + \frac{105}{32}g^2(1-x)^4\right].\label{e40}
\end{eqnarray}

    Exact function $I(g)$ does not depend on $z$ (or $x$) accordingly to (\ref{e30}) or (\ref{e35}), therefore $\frac{\partial I(g)}{\partial x} \equiv 0$. However, any finite series approximation $I_{n}(g,x)$ actually explicitly depends on $x$. According to  the PMS \cite{Stev81}, let us require minimal sensitivity to $x$ as early as in the first order:

\begin{equation}\label{e41}
  \frac{\partial I_{1}(g,x)}{\partial x} = 0 \, ,
\end{equation}
that gives the relation
\begin{equation}\label{e42}
    g = \frac{2}{5} \,\frac{x}{(1-x)^2} \, .
\end{equation}
We see from here  that $0 \leq x \leq 1$ owing to $0 \leq g \leq \infty$. Inverting this equation results in
\begin{equation}\label{e43}
  x(g) = \frac{1+5g-\sqrt{1+10 g}}{5g} \, .
\end{equation}
Substituting (\ref{e42}) into (\ref{e40}), we obtain
\begin{equation}\label{e44}
  I_{2}(x) = \sqrt{1-x}\left[1+\frac{1}{5} x + \frac{3}{20} x^2\right] .
\end{equation}
Finally, after substituting RG approximation (\ref{e17}) instead of the square brackets in (\ref{e44}), we get
\begin{equation}\label{e45}
  I^{(2)}_{\beta}(g) = \sqrt{1-x(g)}\left\{1+\frac{2}{15} \left[\exp\left(\frac{3}{2} x\left(g\right)\right)-1\right]\right\},
\end{equation}
where $x(g)$ is given by formula (\ref{e43}). In accordance with  the general property of the RG approximation method,  the weak coupling limit ($g \ll 1$) of (\ref{e45}) up to the second order coincides with expansion (\ref{e34}). The strong coupling limit of (\ref{e45}), when $g \gg 1$, is
\begin{equation}\label{46}
  I^{(2)}_{\beta}(g) \simeq 1.164\,456\, g^{-1/4}-.634\,951\, g^{-3/4}\, ; \qquad  g \gg 1 \, .
\end{equation}
The accuracy  of the approximation (\ref{e45}) is the monotone increasing function of  $g$ that tends to limit  $\approx 13.8\%$ when $g \to \infty$.

\subsection{Improved modified RG approximation.}\label{c:IM RG PF}

It is important to note, that relation (\ref{e42}) was obtained by
applying the PMS to the first order approximation (\ref{e39}) but
not to the RG approximation.  Therefore the relation (\ref{e42})
is not optimum for the RG approximation to achieve the best
approximation of the exact function $I(g)$. Let us use more
general relation instead of (\ref{e42})
\begin{equation}\label{e47}
    g = \frac{2}{5 p}\frac{x}{(1-x)^2} \, .
\end{equation}
with trial parameter $p$. Inverting this equation gives
\begin{equation}\label{e48}
  x(g) = \frac{1+5p g-\sqrt{1+10 p g}}{5p g} \, .
\end{equation}
Using the relation (\ref{e47}) results in the following RG approximation
\begin{equation}\label{e49}
     I^{(2)}_{\beta}(g) =  \sqrt{1-x(g)}\left\{1+\frac{2}{15} \frac{\left[1+\frac{5}{2} (p-1)\right]^2}{1+\frac{5}{2} \left(p-1 \right)^2} \left[\exp\left(\frac{3}{2}\frac{1+\frac{5}{2} \left(p-1 \right)^2}{p \left[1+\frac{5}{2} \left(p-1 \right)\right]} x\left(g\right)\right)-1\right]\right\},
\end{equation}
where $x(g)$ is given by formula (\ref{e48}).

The weak coupling limit of (\ref{e49}) again coincides with the
expansion (\ref{e34}) up to the second order.  As can be shown,
this remarkable property is the consequence the fact that the
modified perturbation potential $\triangle V$ (\ref{e36})
approaches the initial perturbation potential $g\phi^4$ when $x
\to 0$.

On the other hand, the strong coupling limit of (\ref{e49})
depends on $p$. The optimal value of the parameter $p$ can be
found now from the coincidence of the strong coupling limit of
$I^{(2)}_{\beta}(g)$ with this limit of the exact integral $I(g)$
that is given by (\ref{e32c}), i.e.
\begin{equation}\label{e50}
   \lim\limits_{g \to \infty}\frac{I^{(2)}_{\beta}(g)}{I(g)}=1
  =\left(\frac{2}{5 p}\right)^{1/4}\left\{1+\frac{2}{15} \frac{\left[1+\frac{5}{2} (p-1)\right]^2}{1+\frac{5}{2} \left(p-1 \right)^2} \left[\exp\left(\frac{3}{2}\frac{1+\frac{5}{2} \left(p-1 \right)^2}{p \left[1+\frac{5}{2} \left(p-1 \right)\right]} \right)-1\right]\right\}  \bigg/ 1.022\,765 \, .
\end{equation}
Solving the last equality of (\ref{e50}) with respect to $p$ we get
\begin{equation}\label{e51}
  p \approx 1.779\,643 .
\end{equation}
Using this value, we get the following strong coupling limit of the RG approximation (\ref{e49})
\begin{equation}\label{e52}
  I^{(2)}_{\beta}(g) \simeq 1.022\,765\, g^{-1/4}-.343\,514\, g^{-3/4} , \qquad  g \gg 1.
\end{equation}

By numerical calculations, the maximal error for the RG approximation (\ref{e49}) with parameter $p$ given by (\ref{e51}) is found to be $ \approx 0.06\%$ at $g=1.9$. Remind that when we did not optimized $p$ ($p=1$), the accuracy of the method was $13.8\%$. So, by using asymptotic constraint condition (\ref{e50}) we improved the accuracy more than in two hundred times.

\section{Quartic anharmonic oscillator}\label{c:QO}

Now let us check the ability of our method to approximate the ground state energy $E_0$ of the one-dimensional anharmonic oscillator with the Hamiltonian ($\hbar=1$)
\begin{equation}\label{e53}
  H=-\frac{1}{2m}\frac{d^2}{dy^2}+\frac{m\omega^2}{2}y^2+\lambda m^{2}y^4 \, ,
\end{equation}
in which $m$, $\omega$ and $\lambda$ are positive constants, $y\in (-\infty\, , \; \infty)$.

    Actually, a lot of various physical models can be reduced to the Hamiltonian (\ref{e53}).  The close connection of the quartic-oscillator model with the so-called $\phi^4$ model in the quantum field theory is also well acknowledged.

    The Hamiltonian (\ref{e53}) , making the change $y \to \sqrt{m} y$, can be transformed to the form
\begin{equation}\label{e54}
  H=-\frac{1}{2}\frac{d^2}{dy^2}+\frac{\omega^2}{2}y^2+\lambda y^4 \, ,
\end{equation}
which is more convenient for calculations.
    The perturbation series with respect to the coupling constant $\lambda$ diverges for any finite value of this constant \cite{Bender69,Bender73}. The divergence of the series for the dimensionless ground state energy
\begin{equation}\label{e55}
  e(g) \equiv E_0/\omega \qquad  (g\equiv\lambda/\omega^3)
\end{equation}
is so strong that the expansion in powers of $g$ has no sense at $g \sim 1$. This is due to the increase of coefficients near $g^n$ according to $3^{n}n!$ as $n \to \infty$ \cite{Bender73}. The expansion of (\ref{e55}) in the weak-coupling limit \cite{Bender69,Bender73} is
\begin{equation}\label{e56}
  e(g)\simeq \frac{1}{2}+\frac{3}{4}g-\frac{21}{8}g^2+\frac{333}{16}g^3 -\frac{30885}{128}g^4 \, ; \qquad  (g \ll 1) \, .
\end{equation}
The asymptotic behavior of (\ref{e55}) in the strong-coupling limit is given by \cite{Simon70,HioeMontr75,HioeMontr78}
\begin{equation}\label{e57}
  e(g)\simeq  g^{1/3}\left(0.667\,986+0.143\,67 g^{-2/3}-0.008\,8 g^{-4/6}\right); \qquad  (g \gg 1) \, .
\end{equation}

    To estimate our RG approximation, we take an advantage of exact numerical calculations \cite{Bismas73,HioeMontr75,HioeMontr78} for the ground state energy (\ref{e55}) for a wide range of coupling constants.
Retaining three terms of series (\ref{e56}), we get the approximation of (\ref{e55}), the  relative accuracy $\Delta_{T}^{(2)}$ of which is about 270\%  at $g=1$. In turn, RG approximations (\ref{e17}), (\ref{e19}), and (\ref{e20})  that  based on this approximation give us $\Delta_{\beta}^{(2)} \approx 25\%$, $\Delta_{X}^{(2)} \approx -0.5\%$ and $\Delta_{Xcf}^{(2)} \approx 17\%$  respectively. This is rather reasonable accuracy for approximations based only on two first terms of divergent series.

\subsection{Modified RG approximation.}\label{c:M RG QO}

Now we again improve our method to be applicable to arbitrary strong coupling. For doing this we take as a trial approximation the Hamiltonian
\begin{equation}\label{e58}
  H_0=-\frac{1}{2}\frac{d^2}{dy^2}+\frac{{\omega_{0}}^2}{2}y^2 \, ,
\end{equation}
and calculate the ground state energy for (\ref{e55}) using the perturbation theory over $H-H_0=({\omega}^2-{\omega_{0}}^2) y^2/2+\lambda y^4$. In the zeroth order approximation we have $e_0(g,\omega_0)=\omega_{0}/{2 \omega}$. The first approximation for (\ref{e55}) is
\begin{equation}\label{e59}
  e_{1}(g,\omega_0)=\frac{\omega_{0}}{4 \omega}\left[1+ \left(\frac{\omega}{\omega_{0}}\right)^2 +3g \left(\frac{\omega}{\omega_{0}}\right)^3\right] ,
\end{equation}
and the second one is
\begin{equation}\label{e60}
  e_{2}(g,\omega_0)=e_{1}(g)-\frac{\omega_{0}}{16 \omega} \left\{ \left[1- \left(\frac{\omega}{\omega_{0}} \right)^2 -6 g \left( \frac{\omega}{\omega_{0}} \right)^3 \right]^2+ 6 g^2 \left( \frac{\omega}{\omega_{0}} \right)^6 \right\}.
\end{equation}

    Let us make the substitution $x=1-\left(\omega/\omega_{0} \right)^2$. Then expressions (\ref{e59}) and (\ref{e60}) become
\begin{equation}\label{e61}
  e_{1}(g,x) = \frac{1}{2} \left[1-\frac{1}{2} x+\frac{3}{2} (1-x)^{3/2} g\right] \left(1-x \right)^{-1/2} ,
\end{equation}
\begin{equation}\label{e62}
   e_{2}(g,x) = e_{1}(g,x) -\frac{1}{16} \left\{\left[x-6\left(1-x \right)^{3/2}g\right]^2 +6(1-x)^{3}g^2 \right\} \left(1-x\right)^{-1/2}.
\end{equation}

   Applying again the PMS \cite{Stev81} to the first order approximation (\ref{e61})
\begin{equation}\label{e63}
  \frac{\partial e_{1}(g,x)}{\partial x} = 0 \, ,
\end{equation}
we get the following relation
\begin{equation}\label{e64}
    g = \frac{1}{6} \,\frac{x}{(1-x)^{3/2}} \, .
\end{equation}
We see from here  that $0 \leq x \leq 1$ because of $0 \leq g \leq \infty$. Substituting (\ref{e64}) into (\ref{e62}) we get
\begin{equation}\label{e66}
  e_{2}(x) = \frac{1}{2}  \left[1-\frac{1}{4} x - \frac{1}{48} x^2 \right]\left(1-x \right)^{-1/2} .
\end{equation}
At last, replacing the square brackets in (\ref{e66}) by the RG approximation (\ref{e17}), we get
\begin{equation}\label{e67}
  e_{\beta}^{(2)}(g) =\frac{1}{2} \left\{1-\frac{3}{2} \left[\exp\left(\frac{1}{6} x\left(g\right)\right)-1\right] \right\} \left[1-x\left(g\right)\right]^{-1/2} .
\end{equation}
Here $x(g)$ is positive and continuous solution of the equation (\ref{e64}) with respect to $x$:
\begin{equation}\label{e68}
  x(g) = \left\{
  \begin{array}{lr}
1-\frac{3}{4}\left(g_c/g\right)^2\left[2\cos\left(\alpha/3\right)-1\right]^2  , & \qquad  g \leq g_c \\
1-\frac{3}{4}\left(g_c/g\right)^2\left[A_{+}+A_{-}-1\right]^2, & \qquad  g \geq g_c
  \end{array} \right. \, ,
\end{equation}
where
\begin{eqnarray}\label{e68a}
  g_c & = & \frac{1}{9\sqrt{3}}  \, ,\\
  \alpha & = & \arccos\left(2\left[g/g_c\right]^2-1\right) , \nonumber\\
  A_{\pm} & = & \left\{-1+2\left(g/g_c\right)^2 \pm 2\left(g/g_c\right)\left[\left(g/g_c\right)^2 -1\right]^{1/2}\right\}^{1/3}. \nonumber
\end{eqnarray}

 In accordance with the method of the RG approximation, the week coupling limit of (\ref{e67}) coincides with the expansion (\ref{e56}) up to the second order . The strong coupling limit of (\ref{e67}), when $g \gg 1$, is
\begin{equation}\label{e69}
  e_{\beta}^{(2)}(g) \simeq g^{1/3}\left(.661\,395+.148\,035 g^{-2/3}-.010\,255 g^{-4/3}\right); \qquad  (g \gg 1).
\end{equation}
The accuracy of the approximation (\ref{e67})  monotone decreases with  $g$ and approaches a limit  $\approx 0.99\%$ when $g \to \infty$.

\subsection{Improved modified RG approximation.}\label{c:IM RG QO}

To improve the accuracy we use again more general  relation  than (\ref{e64})
\begin{equation}\label{e70}
    g = \frac{1}{6 p} \,\frac{x}{(1-x)^{3/2}} \, .
\end{equation}
    Using this relation results in the following RG approximation
\begin{equation}\label{e71}
   e_{\beta}^{(2)}(g) =\frac{1}{2}  \left\{1-\frac{3}{2}\frac{(2p-1)^2}{1+6(p-1)^2} \left[\exp\left(\frac{1}{6}\frac{1+6(p-1)^2}{ p\left(2p-1\right)} x\left(g\right)\right)-1\right]\right\} \left[1-x\left(g\right)\right]^{-1/2},
\end{equation}
where $x(g)$ is given by the same formula (\ref{e68}) as before but now $g_c=1/(9\sqrt{3}p)$ instead of (\ref{e68a}).
 The weak coupling limit of (\ref{e71}) coincide again with the expansion (\ref{e56}) up to the second order as in the previous section \ref{c:PF} and on the same reasons. On the other hand, the strong coupling limit of (\ref{e71}) depends on $p$. The optimal value of the parameter $p$ can be found now from the coincidence of the strong coupling limit of  $e_{\beta}^{(2)}(g)$ with known exact limit of  $e(g)$ which is given by (\ref{e57}):
\begin{equation}\label{e72}
   \lim\limits_{g \to \infty}\frac{e^{(2)}_{\beta}(g)}{e(g)}=1
    =  \left(\frac{3 p}{4}\right)^{1/3} \left\{1-\frac{3}{2}\frac{(2p-1)^2}{1+6(p-1)^2} \left[\exp\left(\frac{1}{6}\frac{1+6(p-1)^2}{ p\left(2p-1\right)} \right)-1\right]\right\} \bigg/ 0.667\,986 \,   .
\end{equation}
Solving the last equality of (\ref{e72}) with respect to $p$ we get
\begin{equation}\label{e73}
  p \approx 1.472\,032.
\end{equation}
With this value we get the following strong coupling limit of the RG approximation (\ref{e71})
\begin{equation}\label{e74}
  e(g)\simeq  g^{1/3}\left(0.667\,986+0.143\,62 g^{-2/3}-0.008\,6 g^{-4/6}\right) \, ; \qquad  (g \gg 1)
\end{equation}

By numerical calculations, the maximal relative error for the RG approximation (\ref{e71}) with parameter $p$ given by (\ref{e73}) is found to be $\Delta_{\beta}^{(2)} \approx 0.003\,6 \%$ at $g=0.7$. Remind that when we did not optimized $p$ ($p=1$), the accuracy of the method was $0.99\%$. Thus, we improved the accuracy  in about three hundred  times using asymptotic constraint condition (\ref{e72}).

\section{Conclusion}\label{c:C}

The method of RG approximation presented in this paper possesses the following peculiarities. It seems to be logical and simple. It needs minimal information, e. g., only two first terms of perturbation theory. Even in the simplest form the error of its approximation is about $10\%$ at $g=1$ even for a sought function that has divergent series  in powers of $g$.
 The accuracy of the modified RG approximation that uses the PMS to make it applicable for arbitrary values of $g$  is practically the same. The the accuracy of the approximation can be  drastically improved in hundreds times while using the PMS if we take advantage of some  fitting parameter to be defined by additional constraint condition, e.g., by making equal meaning of the sought function and its approximation at $g \to \infty$. So, the accuracy of the RG approximation of partition functions integral (\ref{e30}) is improved in more than two hundred times and is of $0.06\%$. The accuracy of the RG approximation of the ground state energy of the one-dimensional quartic anharmonic oscillator is also improved in the same way in almost three hundred times and is of $0.003\,6\%$.

The accuracy of our method can be compared with that of other analytical approaches that use an equivalent number of approximate terms, that is with the methods \cite{Stev81,HalliSura80,Yukalov90,Yukalov91,Yukalov93,Yukalov93a}. The first two methods are called the modified perturbation theory, and the others are called the self-similar approximation. All of them also use either the PMS condition similar  to (\ref{e41}), (\ref{e63}) or principle of minimal difference \cite{Yukalov90} to have an opportunity to approximate an interesting physical quantity in the whole range ($g \in{0, \infty}$) of parameter $g$.

  The accuracy of approximation of partition functions integral (\ref{e30}) by  means of both the self-similar approximation \cite{Yukalov93} and the  modified perturbation theory (\ref{e44}) is about $10\%$, practically the same as in our method.  If in the latter one we use  instead of (\ref{e42}) generalized relation (\ref{e47}) with the fitting parameter  $p$ being defined by asymptotic constraint condition similar to (\ref{e50}) then the accuracy of this approximation is improved up to $0.19\%$. As to the self-similar approximation, its accuracy is improved up to $1\%$ by use of the ``fixed-point distance''  fitting parameter  being also defined by  asymptotic constraint condition. Remember that the method of improved modified RG approximation presented in this paper has the accuracy $0.06\%$.

    In the case of calculation of the ground state energy of the one-dimensional anharmonic quartic oscillator the accuracy of the self-similar approximation is about $10\%$ while using the principle of minimal difference \cite{Yukalov90} and is  $0.3\%$ if using the PMS \cite{Yukalov91,Yukalov93a} (the accuracy cited in \cite{Yukalov91} corresponds to only strong-coupling  limit). Different variants \cite{Yukalov76,Yukalov76a,Caswell79,HalliSura80,Stev81,AfanGalp86}  (see also formula (\ref{e66})) of the second-order modified perturbation theory give the ground-state energy with the accuracy not higher than $1\%$. Notice that if we substitute in the second-order approximation (\ref{e62})) instead of (\ref{e64}) the generalized relation (\ref{e70}) with the fitting parameter  $p$ being defined by asymptotic constraint condition similar to (\ref{e72}) then the accuracy of this approximation is improved up to $0.008\%$. Remember that the accuracy of our  modified RG approximation (\ref{e67}) is about $1\%$ whereas the accuracy of improved modified RG approximation (\ref{e71}) is $0.003\,6\%$.

    Our approach allows a number of generalizations.

    In general, the invariance relation (\ref{e8}) with respect to
the RG transformation (\ref{e11}) is valid for any non-degenerate
mapping of an associative set of elements $x$  in other set.
Indeed, relation (\ref{e8}) was obtained with the assumption
that the variable value set $x$  is a semigroup with respect to
`$+$' operation. Evidently this set can be either continuous or
discrete. In the latter case it is possible to represent the
successive approximations of any function $f$  as some discrete
mapping. Let us use the number of the approach $(0,1,2, \ldots)$
as the argument of the discrete mapping and the successive
approximation of the function $(f_0(x), f_1(x), f_2(x), \ldots)$
as the value of the mapping. In this case invariance relation
(\ref{e8}) reads:
\begin{equation}\label{e21}
  F(n+p;f_0(x))=F(n;F(p;f_0(x))),
\end{equation}
where $n$  and $p$  are non-negative integers. This relation
(\ref{e21}) can be rewritten in the other form, if one use these
discrete variables as the subscripts of the mapping $F$:
\begin{equation}\label{e22}
   F_{n+p}(f_0(x))=F_n(F_p(f_0(x))).
\end{equation}
Such functional self-similar transformation was obtained
\cite{Yukalov90} from other reasons and also was used for the
constructing of the so called  self-similar approximation.

Generally a mapping $F$  connects elements $x$  of one set
$M$  of the dimension $m$  and elements $f$  of another set
$N$ of the dimension $n$. In addition the RG invariance relation
(\ref{e8}) is valid in the case of one-to-one correspondence
between the image and preimage of the mapping. For the
continuous sets and differentiable mappings this requirement is
fulfilled, if the rank of the functional determinant $\partial
(f_1, f_2, \ldots, f_n)/\partial  (x_1, x_2, \ldots, x_m)$  is
equal to the least of integers $m$  and $n$  (so called
nondegenerate mapping). In the case of $m < n$  the
nondegenerate mapping $F$  defines some $m-$parametric
nonintersecting subset in the set $N$  (some $m-$dimensional
nonintersecting hypersurface). The particular case $m=1$  and
$n=2$  corresponds to the two-charge quantum-field model
\cite{Bogoliubov55}. In this model the invariant charges obeys
functional equations (\ref{e3}). In case of $m > n$  the mapping
definition domain that is invariant with respect to RG
transformation can be anyone
$n-$dimensional nonintersecting hypersurface in $M$.

    One should notice that the well known monotonous dependence of the invariant charge on the 4-momentum transfer squared in the gauge theories of the electro-weak and strong interactions that are renormalizable seems natural in the light stated above.

    In this paper the RG approximation presented is based on the  expansion of some function $f(x)$ in natural number powers of $(x-x_0)$ around some point $x_0$. However, our approach can be generalized to the case of both meromorphic functions that allow their expansion in positive and negative integer powers and arbitrary (nonmonotonic) functions. The latter generalization consists in  representing of a real function as the real or imaginary part of some complex one-to-one function and  RG approximating of the latter one.

    These and others generalizations of our approach will be given in further publications.


    The author would like to gratefully acknowledge the useful discussion with S.G. Rautian.

\bibliographystyle{utphys_arXiv_hyperlinks}
\bibliography{Nikolaev}
\end{document}